\documentclass[aps, prl, superscriptaddress, twocolumn]{revtex4}
\usepackage{graphicx}

\begin{document}

\title{Sub-harmonic resonant excitation of confined acoustic modes at GHz frequencies with a high-repetition-rate femtosecond laser}

\author{A.~Bruchhausen}
\email{Axel.Bruchhausen@uni-konstanz.de}
\altaffiliation[Also at: ]{Instituto Balseiro \& Centro At\'omico Bariloche (CNEA), and CONICET, Argentina}  
\affiliation{Department of Physics \& Center for Applied Photonics, Universit\"at Konstanz, Germany.}
\author{R.~Gebs}
\affiliation{Department of Physics \& Center for Applied Photonics, Universit\"at Konstanz, Germany.}
\author{F.~Hudert}
\affiliation{Department of Physics \& Center for Applied Photonics, Universit\"at Konstanz, Germany.}
\author{D.~Issenmann}
\affiliation{Department of Physics \& Center for Applied Photonics, Universit\"at Konstanz, Germany.}
\author{G.~Klatt}
\affiliation{Department of Physics \& Center for Applied Photonics, Universit\"at Konstanz, Germany.}
\author{A.~Bartels}
\affiliation{Department of Physics \& Center for Applied Photonics, Universit\"at Konstanz, Germany.}
\author{O.~Schecker}
\affiliation{Department of Physics \& Center for Applied Photonics, Universit\"at Konstanz, Germany.}
\author{R.~Waitz}
\affiliation{Department of Physics \& Center for Applied Photonics, Universit\"at Konstanz, Germany.}
\author{A.~Erbe}
\affiliation{Nanostructures Division, Forschungszentrum Dresden-Rossendorf, Germany.}
\author{E.~Scheer}
\affiliation{Department of Physics \& Center for Applied Photonics, Universit\"at Konstanz, Germany.}
\author{J.-R.~Huntzinger}
\affiliation{CGES-UMR 5650, CNRS, Universit\'e Montpellier 2, France.}
\author{A.~Mlayah}
\affiliation{Centre d'Elaboration de Mat\'eriaux et d'Etudes Structurales CEMES-CNRS, Universit\'e de Toulouse, France.}
\author{T.~Dekorsy}
\affiliation{Department of Physics \& Center for Applied Photonics, Universit\"at Konstanz, Germany.}

\date{\small \today}

\begin{abstract}
We propose sub-harmonic resonant optical excitation with femtosecond lasers as a new method for the characterization of phononic and nanomechanical systems in the gigahertz to terahertz frequency range. This method is applied for the investigation of confined acoustic modes in a free-standing semiconductor membrane. By tuning the repetition rate of a femtosecond laser through a sub-harmonic of a mechanical resonance we amplify the mechanical amplitude, directly measure the linewidth with megahertz resolution, infer the lifetime of the coherently excited vibrational states, accurately determine the system's quality factor, and determine the amplitude of the mechanical motion with femtometer resolution.

\end{abstract}

\maketitle

In recent years, nanophononic and nanomechanical systems have emerged as intriguing subjects for studying mechanics, heat transfer and opto-mechanical coupling on a nanometer scale \cite{Cahill-JApplPhys93-793(03), Kippenberg-Nature456-458(08), Lin-NatPhot4-236(10)}. From a fundamental point of view, they provide a route to study mechanical excitations and their interactions with other elementary excitations \cite{Cahill-JApplPhys93-793(03), Baladin}. >From  an applied perspective they have opened a pathway for high sensitivity sensors in the zeptogram mass range and in the attonewton force range \cite{Yang-NanoLett6-583(06), Mamin-79-3358(01)}. In established experimental methods these systems are driven electrically, magnetically, thermoelastically \cite{Bargatin-APL90-093116(07)}, via radiation pressure from continuous wave lasers \cite{Kippenberg-PRL95-033901(05)}, or via other optical non-radiation-pressure-based schemes \cite{Zalalutdinov-APL79-695(01), Stokes-SensActA21-369(90)}. The frequencies of typical systems investigated so far are in the megahertz to gigahertz frequency range \cite{Bargatin-APL90-093116(07)}. The investigation of higher frequencies is strongly restricted by the driving and detection methods. Here, we report a new method for the investigation of a vibrational system by sub-harmonic resonant excitation with a high-repetition rate femtosecond laser. This excitation scheme can be regarded as tuning the separation of modes of the frequency comb of a femtosecond laser \cite{Udem-Nature416-233(02)} to a commensurable of the frequency of the phononic system. By sweeping the comb spacing of the femtosecond laser, resonant impulsive excitation of the mechanical oscillator can be achieved, which allows the determination of its quality factor in the gigahertz to terahertz frequency range with femtometer sensitivity for the mechanical amplitude. We demonstrate the amplification of the fundamental eigenmode of a free-standing silicon membrane at $19\,\text{GHz}$ by a factor of 20 compared to the off-resonant case and determine its quality factor. \\

The dynamical properties of the free-standing silicon membranes were investigated by performing fs resolution pump-probe experiments using the recently developed high-speed asynchronous optical sampling (ASOPS) method \cite{Bartels-RevScInstr78-035107(07), Hudert-PRB79-201307(R)(09)}. This method is based on two asynchronously linked femtosecond Ti:sapphire ring lasers of repetition rate $f_{_R}\sim 1\,\text{GHz}$. One laser provides the pump beam and the second laser the probe beam. In this technique the time delay between pump-/probe-pulse pairs of the two pulse trains is realized through an actively stabilized $10\,\text{kHz}$ repetition-rate-offset $\Delta f_{_R}$ between the two lasers. This allows the stroboscopic stretching of the time between two consecutive pump pulses ($1\,\text{ns}$) to $100\,\mu\text{s}$ with an increment in time delay between pump and probe pulses of $10\,\text{fs}$. The time resolved reflectivity change of the investigated system is recorded with a detector of $\sim 100\,\text{MHz}$ bandwidth. The absence of mechanical moving parts in this technique in combination with the extremely high scanning rate $\Delta f_{_R}$ allows obtaining a signal-to-noise ratio above $10^7$ in a characteristic acquisition time of a few seconds \cite{Bartels-RevScInstr78-035107(07), Gebs-OptExpress18-5974(10)}. The limit in the determination of the dephasing times of an impulsively driven mechanical system is the available time window of $1\,\text{ns}$ between two successive pump pulses. For any coherent vibrational excitation with a dephasing time larger than the time-delay between two pump pulses, the dephasing time cannot be determined accurately in the time domain. Here we show how this limitation is circumvented by sub-harmonic resonant excitation of the phononic system with a frequency resolution achieved which would require a mechanical delay line of $150\,\text{m}$. 

The pump-probe experiments were performed at room temperature in reflection geometry \cite{Bartels-RevScInstr78-035107(07)}. An average pump and probe power of $100\,\text{mW}$ and $10\,\text{mW}$, respectively, is focused onto a spot of about $50\,\mu\text{m}$ in diameter. The pump and probe wavelengths were set to $825\,\text{nm}$ with a pulse length of $70\,\text{fs}$. The investigated free-standing membranes were prepared from a standard [100]-oriented silicon-on-insulator (SOI) wafer, which was produced via the smart-cut process \cite{Bruel-JpnJApplPhys36-1636(97)}. The membranes were obtained by steps of masking, thermal oxidizing, and etching \cite{Hudert-PRB79-201307(R)(09)}. The single crystalline free-standing membranes were rectangular, roughly $1\times 1\,\text{mm}^2$ in size, and $222\,\text{nm}$ thick. \\

\begin{figure}[t]
\begin{center}
\includegraphics*[keepaspectratio=true, clip=true, angle=0, width=0.65\columnwidth]{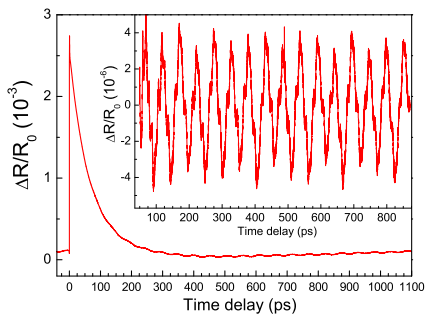}
\caption{(Color online) Time-resolved modulation of the reflected probe beam $\Delta R/R_0$ measured with a repetition rate $f_{_R}=800\,\text{MHz}$. The inset shows these oscillations after subtracting the electronic contribution (Note the change of three orders of magnitude in the $y$-axis between main figure and inset.)}\label{Fig2}
\end{center}
\end{figure}

First we discuss the non-resonant excitation of the membrane, i.e. when the laser repetition rate is not commensurable with the fundamental acoustic mode of the membrane. The acoustic phonon vibrational spectrum of thin free-standing membranes is characterized by equidistant peaks that appear as a consequence of the confinement of the acoustic modes \cite{SotomayorTorres-PhysStatSol(c)1-2609(04), Groenen-PRB77-045420(08), Hudert-PRB79-201307(R)(09)}. In the time domain, this behavior is evidenced by the vibrational superposition of a fundamental mode and oscillations corresponding to odd higher harmonics \cite{Hudert-PRB79-201307(R)(09)}. A transient from a pump-probe experiment taken with a non-resonant repetition rate of $f_{_R}=800\,\text{MHz}$ is shown in Fig.\ref{Fig2}.

\begin{figure}[t]
\begin{center}
\includegraphics*[keepaspectratio=true, clip=true, angle=0, width=\columnwidth]{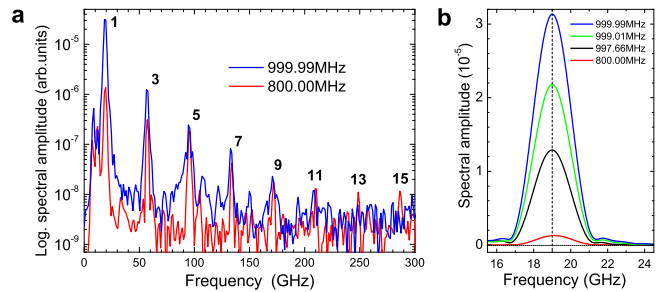}
\caption{(Color online) (a) Fourier spectra of the transients recorded with two different repetition rates. The red curve represents the out-of-resonance situation ($f_{_R}=800\,\text{MHz}$), while the blue curve represents the exact resonant condition 
($f_{_R}=999.99\,\text{MHz}$).  Odd confined modes are labeled with their mode index ($n=1,3,\dots 15$). 
(b) Resonant behavior of the amplitude of fundamental mode $f_n=19\,\text{GHz}$ ($n=1$) when $f_{_R}$ is varied as indicated.}\label{Fig3}
\end{center}
\end{figure}

The transient showing the temporal modulation of the reflected probe beam $\Delta R/R_0$ is characterized by a sharp rise of the signal at $t=0\,\text{ps}$ due to the electronic excitation followed by an exponential decay. The features of interest for this work are the small oscillations that modulate the signal. The inset (in Fig.\ref{Fig2}) shows these extracted oscillations after the subtraction of the electronic contribution which can be modelled by a decaying multiple exponential function. The numeric fast Fourier transform (FFT) of the extracted oscillation clearly exhibits the narrow equidistant peaks corresponding to the longitudinal acoustic phonon modes that are confined perpendicular to the membrane's surface [Fig.\ref{Fig3}a (red curve)]. Due to selection rules in the generation and detection processes, only the odd confined modes are observed \cite{Hudert-PRB79-201307(R)(09)}. The frequency of these modes is given by $f_n=n \frac{v_{ac}}{2d}$ ($n=1,3,5\dots$), where $v_{ac}=8433\,\text{m/s}$ \cite{Groenen-PRB77-045420(08)} is the longitudinal bulk acoustic velocity and $d$ is the thickness of the membrane. For this membrane of $d=222\,\text{nm}$, the fundamental frequency corresponds to $f_1=19\,\text{GHz}$ \cite{Hudert-PRB79-201307(R)(09)}, and modes up to $n=15$ ($285\,\text{GHz}$) are observed.  

\begin{figure}[t]
\begin{center}
\includegraphics*[keepaspectratio=true, clip=true, angle=0, width=0.70\columnwidth]{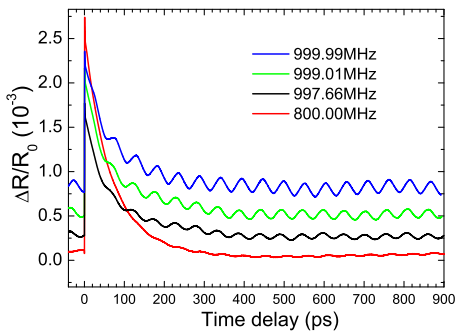}
\caption{(Color online) Typical transients for different pump beam repetition rates ($f_{_R}$) as indicated. $f_{_R}=800\,\text{MHz}$ corresponds to the far-from-resonance situation (also shown in Fig.\ref{Fig2}), and $f_{_R}=999.99\,\text{MHz}$ to the exact resonant condition with the 18$^{th}$ sub-harmonic of the membrane's fundamental confined acoustic mode $f_1=19\,\text{GHz}$. The transients are shown vertically shifted for clarity.}\label{Fig4}
\end{center}
\end{figure}

\begin{figure}[t]
\begin{center}
\includegraphics*[keepaspectratio=true, clip=true, angle=0, width=0.75\columnwidth]{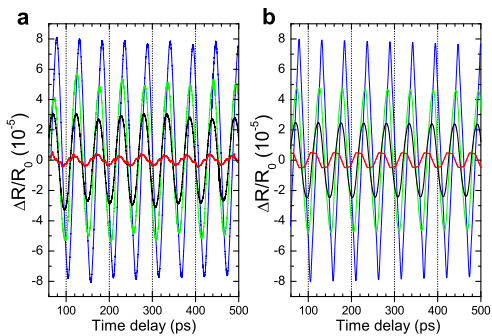}
\caption{(Color online) Oscillations corresponding to the membrane's confined longitudinal acoustic modes for different excitation repetition rates $f_{R}$. (a) Experimantal oscillations, derived from Fig.\ref{Fig4}. (b) Corresponding simulated transients. The color legends correspond to the one shown in Fig.\ref{Fig4}.}\label{Fig5}
\end{center}
\end{figure}

The amplitude of the coherent oscillations barely decays within the time-window of observation of $\sim 1\,\text{ns}$, i.e. the characteristic lifetime ($\tau$) of the phonon vibrations is much longer than $1\,\text{ns}$. At the arrival time of the next pump pulse the phonon system is still coherently excited. In order to impulsively excite acoustic modes constructively with subsequent pump pulses, we adjust the repetition rate of the pump laser to match the phonon oscillations. The resonant excitation of the fundamental mode is carried out by tuning the repetition rate of $f_{_R}\sim 1\,\text{GHz}$ exactly to the 18$^{th}$ sub-harmonic of the membrane's fundamental mode, i.e. $f_{_R}=f_{1}/19$. In resonance, every 19$^{th}$ tooth of the pump laser comb matches one tooth of the comb of the membrane's phonon mode. In the time domain, in this situation the phonon mode is excited by the pump pulses once every 19$^{th}$ oscillating cycle of the phonon vibration, hence the modes are amplified resonantly. This resonance enhancement can be clearly observed in Fig.\ref{Fig4}.
The red and green curves show the close-to-resonance situation, whereas the blue curve represents the exact resonance situation. The corresponding extracted acoustic oscillations are shown in Fig.\ref{Fig5}. The strong amplitude change of the generated reflection modulations is clearly visible, as closer the pump repetition rate approaches the 18$^{th}$ sub-harmonic of $f_1$. In addition a phase shift is observed when $f_{_R}$ is varied. 

Figure \ref{Fig3} shows the Fourier transform of the extracted oscillations depicted in Fig.\ref{Fig5}a. Panel (a) of Fig.\ref{Fig3} compares the two extreme situations: the far-from-resonant situation ($f_{_R}=800\,\text{MHz}$, red curve), and the blue curve corresponding to the exact resonant condition ($f_{_R}=999.99\,{MHz}$). Both curves show the intense and narrow peaks at the modes frequency $f_n$ ($n=1$ to $15$). An increase of about a factor of 20 is observed for the $n=1$ phonon mode. The progressive increase in spectral amplitude of this mode is shown for different repetition rates close to resonance in Fig.\ref{Fig3}b. The width of all peaks in this figure is limited by the experimental time window of observation of $1\,\text{ns}$. 

\begin{figure}[b!!]
\begin{center}
\includegraphics*[keepaspectratio=true, clip=true, angle=0, width=0.60\columnwidth]{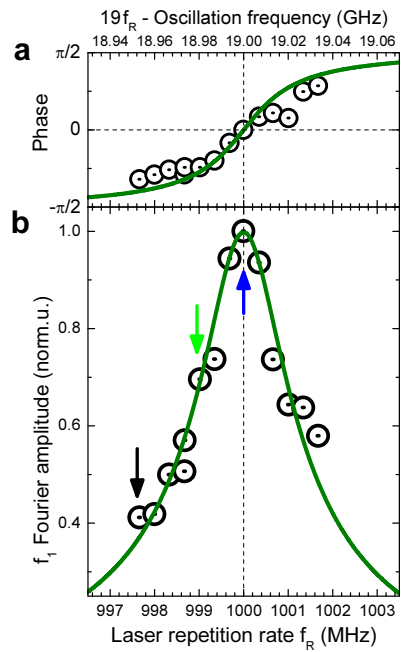}
\caption{(Color online) (a) Phase of the extracted oscillations with respect to the resonant condition, and (b) normalized spectral amplitude of the $f_1$ acoustic confined mode as function of the laser repetition rate $f_{_R}$. The full lines are the result from the simulations using the theoretical model described in the text. The arrows indicate the corresponding $f_{_R}$ positions of the transients and Fourier spectra shown in the Figs.\ref{Fig3} to \ref{Fig5} for the respective colors. The top axis indicates the frequency of the driven $\Delta R/R_0$ oscillations, i.e. 19 times $f_{_R}$.}\label{Fig7}
\end{center}
\end{figure}

The resonant excitation of the fundamental $f_1$ acoustic mode of the membrane is illustrated in Fig.\ref{Fig7}b. The evolution of the spectral Fourier amplitude (see Fig.\ref{Fig3}b) corresponding to this mode is plotted as function of the excitation repetition rate $f_{_R}$ (bottom axis), which is scanned between $997.6\,\text{MHz}$ and $1001.6\,\text{MHz}$.

The amplitude of the mode gets enhanced showing a Lorentzian shaped curve with its maximum exactly at $f_{_R}=f_1/19=1\,\text{GHz}$. The maximum amplitude of the membrane oscillations of $\Delta R/R_0=1.6 \times 10^{-4}$ corresponds to a change in the thickness of the membrane of ~$1.6\,\text{pm}$ which is measured with $1\,\text{fm}$ resolution for a noise limit of $\Delta R/R_0 = 10^{-7}$. Furthermore, from the extracted oscillations in the time domain (see Fig.\ref{Fig5}a) the phase shift of the oscillations relative to the resonance condition can be derived. The evolution of this relative phase follows the typical phase progression of a driven oscillator (Fig.\ref{Fig7}a). \\

In order to obtain a thorough understanding of the membranes dynamics, we compared the experimental results to calculations performed using a detailed model for the optical generation and detection processes, based on a macroscopic elastic and electromagnetic theory \cite{Pu-PRB72-115428(05), Hudert-PRB79-201307(R)(09), Wright-APL66-1190(95)}. 

The large penetration depth of the exciting laser pulse ($\alpha^{-1}\sim8\mu \text{m}\gg d$) \cite{Adachi-PRB38-12966(88)} compared to the fast diffusion of the excited carriers through the thin membrane leads to a very fast homogenization of the carrier distribution. Hence, the optically generated stress can be considered as spatially homogeneous on our experimental time scale ($\sim 1\,\text{ns}$) \cite{Hudert-PRB79-201307(R)(09)}. In a similar way, since the excitation pulse duration ($70\,\text{fs}$) is much shorter than the typical electron-hole pair lifetime in silicon ($10-100\,\mu\text{s}$) \cite{Tyagi-SolStElectr26-577(83)}, the temporal profile of the optically generated stress after one excitation can be described as a Heaviside function. The periodic excitation is described by additionally multiplying the response in the Fourier domain by a Dirac comb of frequency $f_{_R}$. Consequently the reflectivity change induced by the confined vibrations is obtained \cite{Hudert-PRB79-201307(R)(09)}.

In the simulations, the phonon damping constant $\gamma$ \cite{Pu-PRB72-115428(05)} is assumed to be the same for all confined modes and is left free to adjust to the experimental data. The results for the calculated sub-harmonic resonant scan over the fundamental harmonic mode (n=1) are shown in Fig.\ref{Fig7}. The simulated Fourier phase and amplitude for varying excitation repetition rate ($f_{_R}$) are plotted in panel (a) and (b) respectively. The best agreement with the experiment is obtained for $\gamma=0.11\times 10^9\,\text{rad/s}$. This yields a full width at half maximum (FWHM) of the fitted curve of $\Delta f_o=3.2\,\text{MHz}$ at the central frequency $f_{o}=1\,\text{GHz}$, as depicted in Fig.\ref{Fig7}(b). Since the mechanical resonance is at 19 times this frequency, $f_1=19\,f_o$, its FWHM is consequently $\Delta f_1=19\,\Delta f_o=2\sqrt{3}\,\frac{1}{2\pi}\,\gamma=60.8\,\text{MHz}$. 

The experimental results have also been compared to simulations in the time domain. This is shown in Fig.\ref{Fig5}b. As can be observed, the overall agreement with Fig.\ref{Fig5}a is considerably good for the amplitude, frequency and the phase. For each repetition rate only the absolute amplitude $\Delta R/R_0$ for the resonant condition (blue curve) has been adjusted.  

In analogy to a mechanical system, the quality factor of the single crystalline free-standing silicon membrane can be defined as $Q=f_1/\Delta f_1=313$ for the fundamental mode. Such a value represents a high $Q$-factor for a mechanical oscillator system, taking the frequency in the GHz-regime and room temperature conditions into account. This value for $Q$ is almost twice the one reported for this frequency range and temperature \cite{Daly-PRB80-174112(09)}.

Assuming a homogeneous broadening, leads consequently to a lifetime ($\tau=1/\gamma$) for the acoustic mode of $\tau= 9\,\text{ns}$.
This value is a {\em lower} limit for the modes lifetime, since inhomogeneous contributions like thickness fluctuations are not considered in the analysis.
The value $\tau= 9\,\text{ns}$ is found to be slightly larger but of the same order of magnitude as compared to those reported in earlier works in bulk silicon \cite{Duquesne-PRB68-134205(03),Daly-PRB80-174112(09)} based on picoseond ultrasonic experiments in this frequency range ($\sim 20\,\text{GHz}$). However, it is important to notice that the method presented here to obtain the lifetime is conceptually more direct and accurate. In conventional picosecond ultrasonic experiments a metallic transducer (i.e. a thin metallic layer) is used to launch a hypersound pulse into the material, and the lifetime is indirectly determined by analyzing the relative acoustic attenuation between the reflected acoustic echoes \cite{Duquesne-PRB68-134205(03), Daly-PRB80-174112(09), Hao-PRB63-224301(01)}. By the intrinsic nature of this method, the pulses have a more or less defined central frequency, but they are in general spectrally very broad. In addition the evaluation of the phonon lifetimes requires the knowledge of additional parameters such as the Gr\"uneisen parameter, which contributes to a large uncertainty in the exact numbers. By this sub-harmonic resonant driving method, the linewidth is directly measured without the need of making further assumptions, and in contrast to Refs.\cite{Duquesne-PRB68-134205(03), Daly-PRB80-174112(09), Hao-PRB63-224301(01)} we determine the lifetime of a single mode.

In conclusion, we have characterized a phononic system by sub-harmonic--resonant excitation of phonon modes using a $1\,\text{GHz}$ femtosecond laser system. The accessible spectral resolution of this repetition rate scanning ASOPS based technique lies in the MHz range, which is well below the conventionally possible spectral resolution of measuring a time window of $1\,\text{ns}$.
Such a frequency resolution is difficult to achieve with conventional pump-probe techniques, since $1\,\text{MHz}$ resolution would require a mechanical scanning of a time delay of $1\,\mu\text{s}$ corresponding to $150\,\text{m}$. The quality factor $Q$ of the investigated membranes was determined to be $313$ at room temperature. We have established a lower limit for the lifetime of the membrane's fundamental confined acoustic mode of $9\,\text{ns}$.  
This new characterization method enables a thorough investigation of the mechanical, thermal and elastic properties of thin membranes, which might be beneficial for the understanding and development of future nanomechanical and optomechanical systems. The coherent sub-harmonic resonant diving might also provide an alternative, interesting and effective way to achieve coherent generation and amplification of hypersound in terahertz opto-acoustic devices \cite{Trigo-PRL89-227402(02), Beardsley-PRL104-085501(10)}, or as well to amplify spin coherence in semiconductor nanostructures \cite{Wolf-Science294-1488(01)}.\\

A. Bruchhausen thanks the Alexander von Humboldt Foundation (Bonn, Germany) for financial support. This work is supported by the Deutsche Forschungsgemeinschaft (DFG) through the SFB 767, by the Ministry of Science, Research and Arts of Baden-W\"urttemberg (Germany), and by the C'Nano GSO program (France).




\end{document}